\documentclass[twocolumn,showpacs,showkeys,preprintnumbers,amsmath,amssymb]{revtex4}
\usepackage[english]{babel}
\usepackage{graphicx}
\usepackage{bm}
\usepackage{hhline}
\usepackage{wasysym}
\usepackage{amscd}

\newcommand{\C}{\mathbb{C}}
\newcommand{\R}{\mathbb{R}}
\newcommand{\Z}{\mathbb{Z}}
\newcommand{\bsigma}{\bm{\sigma}}
\newcommand{\br}{\mathbf{r}}
\newcommand{\ba}{\mathbf{a}}
\newcommand{\bx}{\mathbf{x}}
\newcommand{\by}{\mathbf{y}}
\newcommand{\bz}{\mathbf{z}}
\newcommand{\bn}{\mathbf{n}}
\newcommand{\bfm}{\mathbf{m}}
\newcommand{\bs}{\mathbf{s}}

\DeclareMathOperator{\Tr}{Tr}
\DeclareMathOperator{\grad}{grad}

\begin{document}
\title{Maximal qubit tomography\footnote{Supported by the Hungarian Research Grant OTKA--F043749.}}
\author{Tam\'as Tasn\'adi}
\affiliation{Department for Mathematical Analysis,\\
Budapest University of Technology and Economics,\\
H-1521 Budapest, POB~91, Hungary.}
\date{March 13, 2008}
\begin{abstract}
A new qubit tomography protocol is introduced, based on a \emph{continuous} positive operator valued measure, which is supported by the set of pure states, and \emph{equivariant} under the symmetry group $SO(3,\R)$ of the qubit state space. Thus the sample data for the tomography protocol is a discrete set of points on the Bloch sphere obtained by a series of independent measurements on identical replicas of the qubit. Although the maximum likelihood estimator uniquely exists, it cannot be explicitly calculated. Instead, we present another convenient, simple estimator, show that it is unbiased and calculate its variance.

Furthermore, a detailed account is given about known properties of related discrete qubit state estimation protocols (like ``minimal qubit tomography''), and the results are compared, discussed.
\end{abstract}
\pacs{03.67.-a, 03.65.Wj, 03.65.Ta}
\keywords{qubit, quantum state estimation, POVM.}
\maketitle

\section{Introduction}
%=====================

The main aim of this paper is to introduce a \emph{continuous} positive operator valued measure (POVM) \cite{NCh:00,AlFa:01}, which is \emph{equivariant} under the symmetry group $SO(3,\R)$ of the qubit state space, and investigate the properties of the qubit state estimation protocol based on the generalized measurement defined by this POVM. There are several similar qubit state tomography protocols in the literature \cite{PHSzSz:06,PHM:07} operating with cleverly chosen \emph{discrete} measurements, which respect only a restricted, discrete group of symmetries. Probably the most famous of them is the ``minimal qubit tomography'' protocol \cite{MinQubit:04}. I found it preferable to be a bit didactic here and to insert a brief account on these discrete qubit tomography protocols, for better comparability of the methods and results, and for the convenience of the reader. Thus the second section together with this introduction summarizes known results, while the new investigations are put in section~\ref{s:contmeas}. The properties of the different protocols are compared in the concluding section.

In quantum information theory \cite{NCh:00} the term \emph{quantum bit} (or \emph{qubit}) refers to the simplest nontrivial quantummechanical system, which has only two independent pure states. The Hilbert space of the qubit is $\C^2$, and its possible states are described by the $2\times 2$ density matrices $\rho$, which can be decomposed in terms of the three Pauli matrices
\begin{equation}\label{e:Pauli}
\begin{aligned}
\sigma_x &= \begin{bmatrix} 0 &1 \\ 1 &0\end{bmatrix},&
\sigma_y &= \begin{bmatrix} 0 &-i \\ i &0\end{bmatrix},
\\
\sigma_z &= \begin{bmatrix} 1 &0 \\ 0 &-1\end{bmatrix},&
\bsigma &=\begin{bmatrix}\sigma_x\\ \sigma_y\\ \sigma_z\end{bmatrix}
\end{aligned}
\end{equation}
in the following way:
\begin{equation}\label{e:rhor}
\begin{aligned}
\rho(\br)&= \frac{1}{2}(I+\br \bsigma) =\frac{1}{2}
\begin{bmatrix} 1+z& x-iy\\ x+iy& 1-z \end{bmatrix},
\\
\br&= \begin{bmatrix}x\\y\\z\end{bmatrix} \in \R^3,
\qquad\qquad
|\br| =r \le 1.
\end{aligned}
\end{equation}
(The condition $r\le 1$ ensures that $\rho(\br) \ge 0$.) It means that the possible states of the qubit are in one-to-one correspondence with the points of the unit ball in $\R^3$, which is conventionally called the \emph{Bloch ball}. The surface of the ball, the \emph{Bloch sphere} ($r=1$) represents the pure states, i.e., rank one projectors. The mapping $\br \mapsto \rho(\br)$ respects convex combination (or ``averaging''), so
\begin{align}\label{e:rhoav}
\rho(\alpha \br_1 +\beta \br_2 ) &=\alpha \rho(\br_1) + \beta \rho(\br_2);&
\langle \rho(\br) \rangle &=\rho(\langle \br \rangle)
\end{align}
provided that $\alpha +\beta =1$.

The state space of the qubit has a nontrivial symmetry group determined by the automorphism group of its event lattice. In terms of density matrices the symmetry transformations are conjugations by unitaries, while in the Bloch ball picture the symmetries correspond to orthogonal rotations of the Bloch ball. The unitary group $SU(2,\C)$ is a two-fold covering of the orthogonal group $SO(3,\R)$, so they both have the same Lie-algebra. The structure of this Lie-algebra is reflected in the multiplication relations of the Pauli matrices, which we shall use in the sequel:
\begin{equation}\label{e:sigma}
\begin{aligned}
\sigma_x^2 &=\sigma_y^2 =\sigma_z^2 =I,&
\sigma_x \sigma_y &=-\sigma_y \sigma_x =i \sigma_z,\\
\sigma_y \sigma_z &=-\sigma_z \sigma_y =i \sigma_x,&
\sigma_z \sigma_x &=-\sigma_x \sigma_z =i \sigma_y.
\end{aligned}
\end{equation}

The principal problem of quantum state estimation, or quantum tomography is to give an accurate estimation $\rho'$ for an unknown qubit state $\rho_0$ by performing certain quantum measurements on multiple replicas of the unknown quantum bit. In the present approach it is essential that all the replicas are in the same state $\rho_0$, only one measurement is performed on each single replica, and the choice of the measurement does not depend on the previous results. Because of the inherent probabilistic nature of quantum mechanics the results of the measurements as well as the estimation $\rho'$ itself are random variables. The estimation is \emph{unbiased} if $\langle \rho' \rangle =\rho_0$, where $\langle \cdot \rangle$ designates the expectation value taken over the possible outcomes of the measurements. The accuracy of the estimator $\rho'$ is characterized by its variance $\big\langle d^2(\rho',\langle\rho' \rangle) \big\rangle$, where $d$ is an appropriate distance on the set of density matrices. A usual choice is the distance $d^2(\rho_1,\rho_2) =\Tr(\rho_2 -\rho_1)^2$ based on the Hilbert-Schmidt norm for selfadjoint matrices.

It is worth noting that quantum state estimation is a much more delicate  problem than its classical counterpart because of two main reasons: $i)$~in quantum mechanics two quantities usually cannot be measured at the same time with arbitrary high precision; and $ii)$~measurements destroy the original state of the system. That is why in quantum tomography different, carefully chosen protocols exist for the measurements applied for the estimation of the unknown state.

In the second section we revisit three existing protocols and their basic properties. In all cases the measurements have a finite set of possible outcomes. The first protocol is based on von~Neumann (projective) spin measurements in three orthogonal directions, thus there are three different measurements, each having two possible outcomes. In the second protocol these three projective measurements are put together to obtain a (non-projective) measurement with six possible outcomes, based on a positive operator valued measure (POVM). The third one, the so called ``minimal state tomography'' protocol \cite{MinQubit:04} is very much alike the previous protocol, but the number of possible outcomes of the POVM measurement is reduced to four, which is a lower bound in qubit state tomography.

In these three protocols the maximum likelihood method is applied to obtain the estimator $\rho'$. Unfortunately in certain cases the likelihood function may take its maximum value outside the Bloch-ball, what makes the further exact analysis rather complicated \cite{PHM:07}. Disregarding this fact, we show that the ``unrestricted'' estimator is unbiased (for the first and third protocol) or asymptotically unbiased (for the second protocol), and we calculate the variance of the (unrestricted) estimator in all cases. The variance goes to zero as the number $N$ of measurements is increased, what justifies that for large $N$ and mixed states the unrestricted and exact estimators are essentially the same. For pure states, however, the unrestricted estimator is not a good choice. Another principal disadvantage of these discrete protocols is the fact that they do not respect the whole symmetry group of the qubit.

This last defect is rectified in the third section, where we present a new protocol for qubit tomography based on a POVM, which is supported on the Bloch sphere, and \emph{equivariant} under the symmetry group of the state space. An interesting novelty is that the POVM applied here is \emph{continuous}, i.e., the corresponding measurement has an infinite number of possible outcomes, namely all the pure states. Although the maximum likelihood estimator cannot be explicitly constructed, we present another simple unbiased estimator and calculate its variance.

We conclude by comparing the results obtained for the different discrete and continuous qubit state estimation protocols.

\section{Discrete measurements}
%==============================
Originally von Neumann defined quantum measurement as choosing one out of a set of pairwise complementary events which form a complete system \cite{JvN:55}. Translating it into an algebraic language, a von~Neumann type (or projective) measurement is defined by a complete set of orthogonal projections $\{ P_s \}_{s\in S}$. Here $S$ is an appropriate index set, and the projections satisfy the relations
\begin{align}\label{e:vNm}
\sum_{s\in S} P_s &=I,&
P_s &=P_s^*,&
P_s P_r &=\delta_{s,r} P_s.
\end{align}
In the state $\rho$ the probability that the measurement results in the event $P_s$ is $p_s =\Tr (P_s \rho)$. Usually it is convenient to ``label'' the possible outcomes, i.e., the index set $S$ by real numbers $a_s$ and define a selfadjoint operator $A= \sum_{s\in S} a_s P_s$ by its spectral decomposition. In this case we interpret $A$ as an \emph{observable}, the $a_s$'s are the possible values of the observable (measurement), and the expectation value of $A$ is given by the well known formula $\Tr (A\rho)$. But the essential part of the von~Neumann type measurement is the orthogonal decomposition~\eqref{e:vNm} of unity. Sometimes there is no natural way (or need) for the embedding $S\stackrel{\subset}{\to} \R$. In this case we can still speak about the probabilities $p_s$, which give a classical probability distribution on $S$, but (without further structure on $S$) the expectation value of the measurement has no meaning.

This scheme can be generalized to the so called \emph{positive operator valued measure} (POVM) \cite{NCh:00,AlFa:01}. In this case the identity is decomposed into the sum of (arbitrary) positive operators:
\begin{align}
\sum_{s\in S} Q_s &= I,&
Q_s &\ge 0.
\end{align}
The possible outcomes of the measurement are labelled by the index set $S$, and the probability of the outcome $s$ in the state $\rho$ is $p_s =\Tr(Q_s \rho)$. As in the previous case, the $p_s$'s define a classical probability distribution on $S$.

The POVM measurements are also called \emph{weak measurements}, particularly if the investigated system is coupled to another system and the POVM measurement is obtained from a projective measurement performed on the composite system \cite{NCh:00,AlFa:01}.

In this section we deal with \emph{discrete measurements}, what means that the index set $S$ is finite. The continuous case is conceptually the same, only technically more difficult. We turn to this in section~\ref{s:contmeas}.

\subsection{Orthogonal spin measurements I. --- Projective case}
\begin{figure}
\begin{center}
\includegraphics[scale=0.4]{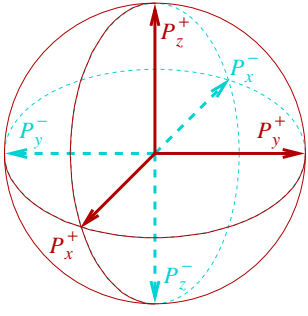}
\end{center}
\caption{\label{f:sp} The six projections belonging to the spin measurements in three orthogonal directions in the Bloch ball.}
\end{figure}

%---------------------------------------------------------------
In this protocol three different von~Neumann spin measurements are performed in three orthogonal directions (see figure~\ref{f:sp}), so the projections are:
\begin{equation}\label{e:Pxyzpm}
\begin{aligned}
P_x^{\pm} &=\rho(\pm\bx)=\frac{I\pm\sigma_x}{2},&
P_y^{\pm} &=\rho(\pm\by)=\frac{I\pm\sigma_y}{2},\\
P_z^{\pm} &=\rho(\pm\bz)=\frac{I\pm\sigma_z}{2},&&
\end{aligned}
\end{equation}
where $\bx$, $\by$ and $\bz$ denote the unit vectors in the three orthogonal directions.

Let $N_x= N_x^+ +N_x^-$, $N_y =N_y^+ +N_y^-$ and $N_z =N_z^+ +N_z^-$ be the number of spin measurements performed in the $x$, $y$ and $z$ direction, where $N_k^+$ is the number of $P_k^+$ results, and $N_k^-$ is the number of $P_k^-$ results ($k= x,y,z$). Furthermore, let $\rho_0 =\rho(\br_0) =\frac{1}{2}(I+\br_0 \bsigma)$ be the unknown true state of the qubit, and let $\rho' =\rho(\br') =\frac{1}{2}(I+\br' \bsigma)$ denote its maximum likelihood estimator.

The probabilities of the different measurement outcomes in the state $\rho_0$ are
\begin{equation}\label{e:ppmxz}
\begin{aligned}
p_x^{\pm} &= \Tr(\rho_0 P_x^{\pm}) =\frac{1\pm x_0}{2},&&
\\
p_y^{\pm} &=\frac{1\pm y_0}{2},&
p_z^{\pm} &=\frac{1\pm z_0}{2},
\end{aligned}
\end{equation}
so the likelihood function $\mathcal L(\br_0)$ for a given measurement statistics $N_k^{\pm}$ is
\begin{multline}\label{e:L}
\mathcal L(\br_0) = 
\binom{N_x}{N_x^+} (p_x^+)^{N_x^+} (p_x^-)^{N_x^-}\\
\times \binom{N_y}{N_y^+} (p_y^+)^{N_y^+} (p_y^-)^{N_y^-}
\binom{N_z}{N_z^+} (p_z^+)^{N_z^+} (p_z^-)^{N_z^-}.
\end{multline}
Varying the argument of the likelihood function, it takes the maximum value either on the boundary of the Bloch ball, or at the zero of its gradient:
\begin{equation}
\grad \mathcal L(\br') = \mathcal L(\br')
\begin{bmatrix}
\frac{N_x^+}{1+x'} -\frac{N_x^-}{1-x'}\\
\frac{N_y^+}{1+y'} -\frac{N_y^-}{1-y'}\\
\frac{N_z^+}{1+z'} -\frac{N_z^-}{1-z'}
\end{bmatrix}
=0
\end{equation}
which yields
\begin{equation}\label{e:rv}
\br' = 
\begin{bmatrix}
\frac{N_x^+ -N_x^-}{N_x}\\
\frac{N_y^+ -N_y^-}{N_y}\\
\frac{N_z^+ -N_z^-}{N_z}
\end{bmatrix}.
\end{equation}

A principal problem with this ``unrestricted'' estimator is that it may fall out of the Bloch ball. In this case either the true maximum place should be found on the boundary of the Bloch ball, or $\br'$ should be normed in some other way, e.g., by dividing it by $|\br'|$. For the moment we disregard this problem and continue as if $\br'$ were a true estimator.

Using the binomial theorem we obtain that
\begin{equation}\label{e:avN+}
\begin{split}
\Big\langle \frac{N^+}{N} \Big\rangle &=
\sum_{N^+ =0}^N \frac{N^+}{N}\binom{N}{N^+} {p^+}^{N^+} {p^-}^{N^-} \\
&= \sum_{N^+ =1}^N \binom{N-1}{N^+ -1} p^+ {p^+}^{N^+ -1} {p^-}^{N^-} =p^+
\end{split}
\end{equation}
in any of the three directions, and applying this and the previous equations \eqref{e:ppmxz}, \eqref{e:L}, \eqref{e:rv} a straightforward calculation shows that the estimator $\br'$ is unbiased:
\begin{equation}
\langle\br'\rangle =\sum_{N_x^+=0}^{N_x} \sum_{N_y^+=0}^{N_y} \sum_{N_z^+=0}^{N_z}
\br' \mathcal L(\br_0) = \br_0.
\end{equation}
(The calculation can be carried out separately in the three spin directions.)

Now let us determine the variance of the unrestricted estimator. Again a simple application of the binomial theorem shows that
\begin{multline}\label{e:avN+2}
\Big\langle \frac{N^+ (N^+-1)}{N (N-1)} \Big\rangle =
\\
=\sum_{N^+ =2}^N \binom{N-2}{N^+ -2} {p^+}^2 {p^+}^{N^+ -2} {p^-}^{N^-} ={p^+}^2
\end{multline}
in any of the three spin directions. The  distance square between the expectation value $\langle \rho' \rangle =\rho_0$ and the estimator is
\begin{equation}\label{e:d2}
d^2(\rho',\langle \rho' \rangle) =\Tr(\rho'-\langle\rho'\rangle)^2 =
\frac{1}{4} \Tr(\Delta\br \bsigma)^2 = \frac{1}{2}\Delta r^2.
\end{equation}
(Here $\Delta \br$ is a shorthand notation for the vector $\br' -\langle \br' \rangle$, and $\Delta r$ is its absolute value.) Again the three directions decouple, what simplifies the calculations. The average of $\Delta x^2$ can be determined with the help of \eqref{e:avN+2}:
\begin{equation}
\langle \Delta x^2 \rangle = \langle x'^2 \rangle -\langle x'\rangle^2 =
\Big\langle \frac{(2N_x^+ -N_x)^2}{N_x^2} \Big\rangle -x_0^2 =
\frac{1-x_0^2}{N_x}.
\end{equation}
It means that in the states $\rho_0 =P_x^{\pm}$ the variance in the $x$ direction is zero, as it is expected. The total variance is obtained by summing similar expressions for the $x$, $y$ and $z$ direction:
\begin{equation}\label{e:V3spin}
\begin{aligned}
V_{\ast} &=
\langle d^2(\rho',\rho_0)\rangle =
\frac{1}{2} \langle \Delta x^2 +\Delta y^2 + \Delta z^2 \rangle
\\
&=\frac{1-x_0^2}{2N_x} +\frac{1-y_0^2}{2N_x} +\frac{1-z_0^2}{2N_z}.
\end{aligned}
\end{equation}

If we assume that $N_x =N_y =N_z =\frac{N}{3}$, where $N =N_x +N_y +N_z$ is the total number of measurements, then the variance has the form
\begin{equation}\label{e:V3}
V_{\ast} = \frac{9-3r_0^2}{2N}.
\end{equation}

The variance tends to zero as the number of measurements goes to infinity, what justifies that for large $N$ and for mixed states the unrestricted estimator \eqref{e:rv} practically coincides with the true maximum likelihood estimator.

\subsection{Orthogonal spin measurements II. --- POVM case \label{ss:hexqubt}}
%-----------------------------------------------------------------------------
This protocol is also based on the six projections $P_{x,y,z}^{\pm}$ appearing in the spectral decomposition of the three orthogonal spin operators (see equation~\eqref{e:Pxyzpm}), but this time we form a single POVM out of them, by decomposing the unity in the following way:
\begin{equation}
I =Q_x^+ +Q_x^- +Q_y^+ +Q_y^- +Q_z^+ +Q_z^- ,
\end{equation}
where $Q_k^{\pm} = \frac{1}{3} P_k^{\pm}$ for $k \in \{x,y,z\}$ [see equation~\eqref{e:Pxyzpm}]. Performing this POVM measurement on each replica of the qubit, the six different results are obtained with the probabilities
\begin{equation}
\begin{aligned}
q_x^{\pm} &=\Tr(\rho_0 Q_x^{\pm}) =\frac{1\pm x_0}{6},&&
\\
q_y^{\pm} &=\frac{1\pm y_0}{6},&
q_z^{\pm} &=\frac{1\pm z_0}{6},
\end{aligned}
\end{equation}
provided that the qubit is in the state $\rho_0 =\rho(\br_0)$.

Let $N_k =N_k^+ +N_k^-$ (for $k\in \{ x,y,z \}$) be the number of measurements with any of the two possible results in the $k$ direction, and let $N=N_x +N_y +N_z$ denote the total number of measurements. The probability of a given measurement statistics $\{ N_k^{\pm} \}$ in the state $\rho_0$ is described by the multinomial distribution:
\begin{equation}
\mathcal L(\br_0) =
N! \prod_{k\in \{x,y,z\}} \frac{(q_k^+)^{N_k^+} (q_k^-)^{N_k^-}}{N_k^+ ! N_k^- !}.
\end{equation}

The unrestricted maximum likelihood estimator $\br'$ is obtained again by determining the zero of the gradient of $\mathcal L(\br')$, and we get formally the same expression~\eqref{e:rv} as in the previous subsection. Now, however, an unavoidable problem is the fact that $N_x$, $N_y$ or $N_z$ may be zero with positive probability! If it happens then we have no information at all about certain component(s) of the Bloch vector $\br'$, and the division in \eqref{e:rv} is meaningless. In this case the most natural thing is to use the estimation $0$ for the appropriate component of $\br'$. (We have to take care of this in the evaluation of the expectation values.) Of course this `ad hoc' choice will bias the estimator towards zero. In addition, the estimator may fall out of the Bloch ball; we disregard it in the followings.

Applying the multinomial theorem it is easy to show that 
\begin{widetext}
\begin{multline}\label{e:avNx+}
\langle x' \rangle =
\Big\langle \frac{N_x^+-N_x^-}{N_x} \Big\rangle_{N_x \ge 1} =
\sum_{\{N_k^{\pm}\}, N_x \ge 1} \frac{N_x^+ -N_x^-}{N_x} \mathcal L(\br_0) 
=\sum_{N_x =1}^N \binom{N}{N_x} (1-q_x)^{N-N_x} \\
\times \sum_{N_x^+ =0}^{N_x} 
\frac{N_x^+ -N_x^-}{N_x} \binom{N_x}{N_x^+} (q_x^+)^{N_x^+} (q_x^-)^{N_x^-}
=\frac{q_x^+ -q_x^-}{q_x} \sum_{N_x =1}^N \binom{N}{N_x} (1-q_x)^{N-N_x} q_x^{N_x}
=x_0 \bigg(1-\Big(\frac{2}{3}\Big)^N \bigg).
\end{multline}
\end{widetext}
(Here $q_x =q_x^+ +q_x^- =\frac{1}{3}$, independently of the true state $\rho_0$.) Similar expressions are valid in the other directions, and it follows that the estimator $\br'$ is biased, but asymptotically unbiased:
\begin{equation}\label{e:rbias}
\langle \br' \rangle =\br_0 \bigg(1-\Big(\frac{2}{3}\Big)^N \bigg).
\end{equation}
Note that the factor $1-(2/3)^N$ is exactly the probability that none of the $N$ measurement results lies in a particular coordinate direction.

To determine the variance of the estimator we need averages of the following type:
\begin{widetext}
\begin{equation}\label{e:avNx+Nx-}
\begin{split}
\Big\langle \frac{N_x^+ N_x^-}{N_x^2} \Big\rangle_{N_x \ge 1} &=
\sum_{N_x =1}^N \binom{N}{N_x} \Big(\frac{2}{3} \Big)^{N-N_x} 
\sum_{N_x^+ =0}^{N_x} \frac{N_x^+ N_x^-}{N_x^2} \binom{N_x}{N_x^+}
(q_x^+)^{N_x^+} (q_x^-)^{N_x^-}\\
&=9q_x^+ q_x^- \sum_{N_x =1}^N \Big( 1-\frac{1}{N_x} \Big) \binom{N}{N_x}
\Big(\frac{2}{3} \Big)^{N-N_x} \Big(\frac{1}{3} \Big)^{N_x}
=\frac{1-x_0^2}{4} \bigg( 1-\Big(\frac{2}{3}\Big)^N -\frac{F_N}{N} \bigg),
\end{split}
\end{equation}
\end{widetext}
where
\begin{equation}\label{e:FN}
\begin{split}
F_N &=N \Big(\frac{2}{3}\Big)^N \sum_{n =1}^{N} \binom{N}{n} \frac{1}{n} 
\Big(\frac{1}{2}\Big)^{n} \\
&=N \Big(\frac{2}{3}\Big)^N \bigg(
\sum_{n =1}^N \frac{1}{n} \Big(\frac{3}{2}\Big)^{n}
-\sum_{n =1}^{N} \frac{1}{n} \bigg).
\end{split}
\end{equation}
Here the last equation was obtained by integrating the following sum with respect to $q$ from $0$ to $\frac{1}{2}$:
\begin{equation}
\frac{d}{dq} \sum_{n =1}^{N} \binom{N}{n} \frac{1}{n} q^{n}=
\frac{(1+q)^N -1}{q} =
\sum_{n=0}^{N-1} (1+q)^n.
\end{equation}
Unfortunately the sum~\eqref{e:FN} defining $F_N$ cannot be further simplified, but it can be shown that $\lim_{N\to \infty} F_N =3$. Indeed, the second sum in \eqref{e:FN} diverges only logarithmically, but it is multiplied with an exponentially descending factor, so it rapidly converges to zero. The first sum $S_N =N \sum_{n=1}^N \frac{1}{n} (2/3)^{N-n}$ satisfies the recursion
\begin{equation}
S_{N+1} -S_N = \Big(\frac{2}{3N} -\frac{1}{3} \Big)S_N +1,
\end{equation}
which means that $\lim_{N\to \infty} S_N =\lim_{N\to\infty} F_N =3$.

Using~\eqref{e:avNx+} and \eqref{e:avNx+Nx-} the average of the error square of the $x$ component of the estimator is
\begin{equation}
\begin{split}
\langle \Delta x^2 \rangle &=
\Big\langle \frac{N_x^2 -4N_x^+ N_x^-}{N_x^2} \Big\rangle_{N_x \ge 1} 
-\langle x' \rangle^2 \\
&= \frac{F_N}{N} (1-x_0^2) +x_0^2 \Big(\frac{2}{3} \Big)^N 
\bigg(1-\Big(\frac{2}{3}\Big)^N \bigg),
\end{split}
\end{equation}
and similar formulas are valid in the other directions. By equation~\eqref{e:d2} the variance of the estimator is:
\begin{equation}\label{e:V6}
V_{\circledast} =\frac{F_N (3-r_0^2)}{2N} +r_0^2 \frac{6^N -4^N}{9^N} \approx
\frac{9-3r_0^2}{2N}.
\end{equation}
For large $N\gg 1$ this expression has the same asymptotic behavior as the variance~\eqref{e:V3} of the estimator using projective spin measurements in orthogonal directions.
\subsection{Minimal qubit tomography \label{ss:minqubt}}
\begin{figure}
\begin{center}
\includegraphics[scale=0.4]{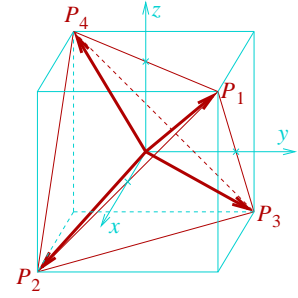}
\end{center}
\caption{\label{f:min} The four projections $P_k$ at the vertexes $\ba_k$ of a regular tetrahedron on the Bloch sphere.}
\end{figure}

%-------------------------------------------------------
This protocol was introduced in \cite{MinQubit:04}. The protocol is based on a POVM, which is minimal in the sense that it contains only four positive operators $\{Q_k\}_{k=1}^4$ in the decomposition of unity. These operators are constant multiples of the four projectors $P_k =\rho(\ba_k)$ being at the vertexes $\{\ba_k\}_{k=1}^4$ of a regular tetrahedron on the Bloch sphere, as shown in figure~\ref{f:min}:
\begin{equation}\label{e:a14}
\begin{aligned}
\ba_1 &=\frac{1}{\sqrt{3}}\begin{bmatrix} 1\\ 1\\ 1\end{bmatrix},&
\ba_2 &=\frac{1}{\sqrt{3}}\begin{bmatrix} 1\\ -1\\ -1\end{bmatrix},
\\
\ba_3 &=\frac{1}{\sqrt{3}}\begin{bmatrix} -1\\ 1\\ -1\end{bmatrix},&
\ba_4 &=\frac{1}{\sqrt{3}}\begin{bmatrix} -1\\ -1\\ 1\end{bmatrix},
\end{aligned}
\end{equation}
\begin{align}
Q_k &= \frac{1}{4} \rho(\ba_k) =\frac{I+\ba_k \bsigma}{8},&
I&= \sum_{k=1}^4 Q_k.
\end{align}

Due to the tetrahedral symmetry, the vectors $\ba_k$ have the following properties:
\begin{align}\label{e:propa}
\ba_k \ba_l &= -\frac{1}{3} +\frac{4}{3} \delta_{k,l}&
\sum_{k=1}^4 \ba_k \circ \ba_k &=\frac{4}{3} I.
\end{align}

In the state $\rho_0 =\rho(\br_0)$ the probabilities $p_k$ of the four different measurement outcomes can easily be calculated with the help of the properties~\eqref{e:sigma} of the Pauli matrices:
\begin{equation}\label{e:p4}
p_k =\Tr(\rho_0 Q_k) =\frac{\Tr\big((I+\br_0 \bsigma) (I+\ba_k \bsigma)\big)}{8} =
\frac{1+\br_0 \ba_k}{4}.
\end{equation}
Denoting the number of $k$-th measurement results by $N_k$ and the total number of measurements by $N =\sum_{k=1}^4 N_k$, the probability of a given measurement statistics in the state $\rho_0$ is described again by the multinomial distribution:
\begin{equation}
\mathcal L(\br_0) =\frac{N!}{N_1! N_2! N_3! N_4!} 
p_1^{N_1} p_2^{N_2} p_3^{N_3} p_4^{N_4}.
\end{equation}
This likelihood function takes its maximal value at the zero of its gradient, which yields the equation
\begin{equation}\label{e:lemin}
\sum_{k=1}^{4} \frac{N_k \ba_k}{1+\br' \ba_k} =0
\end{equation}
for the Bloch vector $\br'$ of the estimator $\rho' =\rho(\br')$. For general $\ba_k$ vectors it is hopeless to solve explicitly this equation, but for the symmetrically distributed vectors~\eqref{e:a14}, using their properties~\eqref{e:propa} it is easy to check that the unique solution is 
\begin{equation}\label{e:rv4}
\br' =3\sum_{k=1}^{4} \frac{N_k}{N} \ba_k.
\end{equation}
This solution is well defined for any measurement statistics $\{ N_k\}$, but it may fall out of the Bloch ball.

Using the multinomial theorem it is easy to show that $\langle N_k \rangle =Np_k$, so
\begin{equation}
\langle \br' \rangle =3\sum_{k=1}^4 p_k \ba_k =
\frac{3}{4} \sum_{k=1}^4 \ba_k + 
\frac{3}{4}\bigg(\sum_{k=1}^4 \ba_k \circ \ba_k \bigg) \br_0 =\br_0,
\end{equation}
which means that the estimator $\rho'$ is unbiased.

Applying again the multinomial theorem, the formulas~\eqref{e:p4} for the probabilities and the properties~\eqref{e:propa} of the $\ba_k$ vectors, we obtain that
\begin{equation}
\begin{split}
\Big\langle \sum_{k=1}^4 N_k^2 \Big\rangle &=
\sum_{k=1}^4 \Big( \big\langle N_k (N_k -1)\big\rangle +\langle N_k \rangle \Big)\\
&=\sum_{k=1}^4 \big( N(N-1)p_k^2 + N p_k \big)\\
&=\frac{3+r_0^2}{12}(N^2 -N) +N.
\end{split}
\end{equation}
From this, using equations~\eqref{e:rv4} and \eqref{e:propa} we immediately get that
\begin{equation}
\big\langle r'^2 \big\rangle =
\frac{12}{N^2} \Big\langle \sum_{k=1}^4 N_k^2 \Big\rangle-3 =
r_0^2 +\frac{9-r_0^2}{N},
\end{equation}
and finally, using~\eqref{e:d2}, for the variance we obtain:
\begin{equation}\label{e:V4}
V_{\otimes} =\frac{\big\langle r'^2 \big\rangle -\langle \br' \rangle^2}{2} =
\frac{9-r_0^2}{2N}.
\end{equation}

Comparing it with the variances~\eqref{e:V3} and \eqref{e:V6} obtained for the previous protocols, we see that in all cases the variances decrease as $1/N$, but for states close to pure states (i.e., if $r_0 \lessapprox 1$) the minimal qubit tomography protocol is less performant by a factor of $\frac{9-1}{9-3}= \frac{4}{3}$.

\section{Continuous, equivariant measurement \label{s:contmeas}}
%===============================================================

Up to this point we have investigated three different qubit tomography protocols, which were based on (either projective or POVM) measurements with only a finite number of possible outcomes. In a finite $d$ dimensional Hilbert space the maximal number of possible outcomes of a projective measurement is $d$, but this restriction does not hold for generalized, POVM measurements. In this section we construct a new qubit tomography protocol based on a \emph{continuous} POVM measurement, which is \emph{equivariant} under the symmetry group of the qubit system, and supported by the set of pure states, i.e., by the Bloch sphere.

The attribute ``continuous'' refers to the fact that the investigated measurement has an infinite number of possible outcomes, which form a topological manifold---the Bloch sphere in our case. The attribute ``equivariant'' refers to a kind of nice symmetry property, which deserves a deeper investigation.

\subsection{Equivariance \label{ss:equiv}}
%-----------------------------------------

At an abstract level the symmetry group $G$ of an (either classical or quantum) system is an automorphism group of its event lattice \cite{BirkNeu:36,JvN:55}. States are probability measures on the event lattice, so $G$ acts both on events and (by pullback construction) on states. In quantum mechanics events are projections in the Hilbert space $\mathcal H$ of the system, and by Gleason's theorem \cite{Gleas:57} states are represented by density operators. By Wigner's theorem~\cite{Wig:31}, if $\dim \mathcal H >2$, every automorphism $g$ of the projector lattice can be realized by conjugation with a unitary (or antiunitary) operator $U_g$, thus we obtain a left $G$-action on the lattice of projections (events) $\mathcal P$, and a right $G$-action on the set of density operators (states) $\mathcal S$:
\begin{align}\label{e:GactP}
G\times \mathcal P &\to \mathcal P.&
(g,P) &\mapsto gP=U_g PU_g^*;
\\ \label{e:GactS}
\mathcal S \times G &\to \mathcal S,&
(\rho,g) &\mapsto \rho g=U_g^* \rho U_g,
\end{align}
and, by the pullback construction
\begin{equation}\label{e:rhogp}
\rho(gP) =\Tr(\rho U_g P U_g^* )= \Tr(U_g^* \rho U_g P) =(\rho g)(P).
\end{equation}

Performing many times a measurement with possible outcomes in the Borel space $(S,\mathcal B)$ we obtain a probability distribution on the space $S$. Thus a measurement can be regarded as assigning to a quantum state $\rho$ (i.e., probability function on $\mathcal P$) a classical state $p_{\rho}$ (i.e., a probability function on $\mathcal B$). The simplest way to do this is pulling back $\rho$ by a morphism $P:\mathcal B \to \mathcal P$ between the classical and quantum event lattice:
\begin{align}
p_{\rho} =\rho \circ P: \mathcal B &\to [0,1], &
E &\mapsto p_{\rho}(E) =\Tr\big(\rho P(E)\big).
\end{align}
The classical $\to$ quantum lattice morphism $P$ is nothing but a projector valued measure, and the above formula is exactly the probability for the outcome of the projective measurement $P$ being in $E$. Formally the generalized, POVM measurements are obtained by changing $P$ to a positive operator valued measure $Q$ on $(S,\mathcal B)$.

Now let us assume that $S$ is a right $G$-space, and let $\gamma_g :S \to S$ denote the transformation corresponding to $g\in G$. (So $\gamma_g \circ \gamma_h =\gamma_{hg}$.) Then its inverse image $\gamma_g^{-1}: \mathcal B \to \mathcal B$ defines a left action of $G$ on $\mathcal B$. We say that the POVM $Q$ is \emph{equivariant} with respect to the group actions $\gamma^{-1}$ and $\Hat{U} =U\cdot U^*$ (conjugation by $U$) if for any $E\in \mathcal B$ and $g\in G$ the following diagram commutes:
\begin{align}\label{e:genimp}
\begin{CD}
\mathcal B @>{\gamma_g^{-1}}>> \mathcal B\\
@VQVV @VQVV \\
B(\mathcal H) @>{\Hat{U}_g}>> B(\mathcal H)
\end{CD}
&&
U_g Q(E) U_g^* &= Q\big( \gamma_g^{-1}(E) \big),
\end{align}
where $B(\mathcal H)$ denotes the set of bounded operators in $\mathcal H$.

For projector valued measures this structure was discovered by George W. Mackey \cite{Mac:76,Mac:78}, while he was studying the foundations of quantum mechanics \cite{Mac:63}, and he called it \emph{imprimitivity system}. Mackey, in his \emph{imprimitivity theorem}, classified and explicitly constructed the possible imprimitivity systems using \emph{induced representations}. Since then induced representation became a very important tool in the theory of unitary group representation and harmonic analysis, while the imprimitivity theorem turned out to be a cornerstone of quantum mechanics of free systems. Indeed, the equivariance condition~\eqref{e:genimp}, stated for projector valued measures on the three dimensional Euclidean space, is a very general and deep expression of the canonical commutation relations. Starting from this, many important properties (like the existence of mass, spin) of the elementary particles can be deduced, based on purely spacetime symmetry requirements \cite{Mac:63,Mac:78,Varad:68,Jau:68}.

Borrowing the terminology from the projective case, we refer to the equivariance condition~\eqref{e:genimp} as \emph{general system of imprimitivity}, more precisely, as the POVM $Q$ is a \emph{generalized system of imprimitivity for $U$ based on $S$}. In the next subsection we give a very simple example for this in terms of a qubit.

\subsection{Equivariant POVM on the Bloch sphere \label{ss:equPOVM}}
%-------------------------------------------------------------------

First of all, we note that Wigner's theorem does not apply for two dimensional Hilbert spaces; the projector lattice of $\C^2$ has a much bigger symmetry group than the symmetries induced by unitaries (and antiunitaries). People consider it as a pathological fact due to the low dimensionality of the space. Here we join this opinion and consider the group $SO(3,\R) =SU(2,\C)/\Z_2$ as the symmetry group $G$ of the qubit. (The reason for the factorization is that conjugation with $U$ and $-U$ is the same transformation.)

Furthermore, as the set $S$ of possible measurement outcomes we choose the set of pure states, i.e., the Bloch sphere itself! It is a very clever choice for several reasons: $i)$~By~\eqref{e:GactS} $S$ is a right $G$-space in a natural way. $ii)$~Classically pure states are Dirac measures on the phase space, thus in this respect $S$ is the quantum analog of classical phase space. Is there any better measurement than the one returning \emph{the} (or \emph{a}?) phase space position (pure state) of the measured system? And finally $iii)$~pure states are rank one projections, i.e., positive operators, thus the POVM $Q$ is ``already there'' on $S$, one has only to normalize it.

The rank one projectors $\rho(\bs)$ are parametrized by points $\bs$ of the Bloch sphere ($|\bs|=1$), unitaries have the form $U_{\bn} =e^{i\bn\bsigma}$ (where $|\bn| < 2\pi$), and it can be shown that
\begin{equation}\label{e:UrhoU}
U_{\bn} \rho(\bs) U_{\bn}^* =
e^{i\bn\bsigma} \rho(\bs) e^{-i\bn\bsigma}=
\rho(O_{-2\bn} \bs),
\end{equation}
where $O_{\bfm}$ is the orthogonal rotation around the axis $\bfm$ at an angle $|\bfm|$. (To derive this the equation $[\bn\bsigma, \bfm \bsigma] =2i(\bn \times \bfm) \bsigma$ is needed, which is a straightforward consequence of~\eqref{e:sigma}.) Thus in the Bloch sphere picture the symmetry transformations are the rotations of the sphere! (We remark here that the nonunitary symmetries are arbitrary homeomorphisms of the Bloch sphere which map antipodal points to antipodal points.) Equations~\eqref{e:GactS} and \eqref{e:UrhoU} also show that in this case the right $G$-action $\gamma_{\bn}$ on the space $S$ is
\begin{equation}\label{e:gam}
\gamma_{\bn}(\bs) =\text{representing point of }U_{\bn}^* \rho(\bs) U_{\bn} =O_{2\bn} \bs.
\end{equation}

Now let $\omega$ be the normalized area measure on the unit sphere $S$, i.e., in spherical polar coordinates $(\vartheta, \varphi)$, $d\omega(\vartheta, \varphi) =\frac{\sin \vartheta}{4\pi} d\vartheta d\varphi$, and for a Borel set $E\subset S$ let the POVM $Q$ be defined by
\begin{subequations}\label{e:Q}
\begin{equation}\label{e:Qa}
Q(E) =2\int_{\bs\in E} \rho(\bs) d\omega(\bs) 
=2\omega(E) \rho\big(
{\textstyle \int_{\bs \in E} \bs d\omega(\bs)}\big),
\end{equation}
or shortly
\begin{align}\label{e:Qb}
dQ(\bs) &=2\rho(\bs) d\omega(\bs),&
Q &= 2\rho \omega. 
\end{align}
\end{subequations}
(The last equality in~\eqref{e:Qa} is a consequence of~\eqref{e:rhoav}.) It is clear that $Q$ is normalized, i.e., $Q(S) =2\rho(\mathbf{0}) =I$.

Using the previous three formulas~(\ref{e:UrhoU}--\ref{e:Q}) and the invariance of the area measure $\omega =\omega\circ O^{-1}$ under any rotation $O$, it is simple to show that $Q$ satisfies the equivariance condition~\eqref{e:genimp}:
\begin{multline}
U_{\bn} Q(E) U_{\bn}^* =
2{\textstyle \omega(E)\rho\big( \int_{\bs \in E} O_{-2\bn} \bs d\omega(\bs) \big)}
\\
=2{\textstyle \omega(E)\rho\big( \int_{\bs \in O_{-2\bn}E} \bs d\omega(\bs) \big)}
=Q\big( \gamma_{\bn}^{-1}(E)\big).
\end{multline}
(It is worth noticing that here essentially the equation~\eqref{e:rhogp} and the invariance of $\omega$ was used.)

We give the concrete formula of the POVM $Q$ in spherical polar coordinates $(\vartheta, \varphi)$, although we will keep on using the abstract form~\eqref{e:Q}:
\begin{equation}
dQ(\vartheta, \varphi) =\frac{\sin\vartheta}{4\pi}
\begin{bmatrix}
1+\cos\vartheta & e^{-i\varphi} \sin\vartheta\\
e^{i\varphi} \sin\vartheta & 1-\cos\vartheta
\end{bmatrix}
 d\vartheta d\varphi.
\end{equation}

Intuitively this POVM can be regarded as the limit of discrete POVM's supported by more and more points scattered uniformly on the Bloch sphere. In the previous section we have seen examples for four (subsection~\ref{ss:minqubt}) and six (subsection~\ref{ss:hexqubt}) supporting points with tetrahedral and hexagonal symmetry, respectively.

Now let us investigate the distribution of the measurement $Q$ provided that the system is in state $\rho_0 =\rho(\br_0)$. The probability density function $f_0$ (with respect to $\omega$) at the outcome $\bs \in S$ is
\begin{equation}\label{e:f0}
\begin{split}
f_0 (\bs) &=\lim_{dE\to 0}\frac{\Tr\big(\rho_0 Q(dE)\big)}{\omega(dE)}\\
&=2\Tr\big(\rho(\br_0)\rho(\bs)\big) 
= 1+\br_0 \bs,
\end{split}
\end{equation}
where $dE$ is an infinitesimally small area on the sphere $S$ around $\bs$. (The last equality is a direct consequence of~\eqref{e:sigma}.) This probability distribution is \emph{unsharp}, even for pure states! If $r_0 =1$, then the maximum is $f_0(\br_0) =2$, and the minimum is $f_0(-\br_0) =0$.

Here it is worth noting that the POVM $Q$ defined in~\eqref{e:Q} is not the unique solution of the equivariance condition \eqref{e:genimp} (but probably the most reasonable one). Indeed, for any $\alpha \in [0,1]$ the POVM
\begin{equation}
Q_{\alpha} (E) =\alpha Q(E) + (1-\alpha)\omega(E) I
\end{equation}
satisfies the equivariance condition~\eqref{e:genimp}, but the density function of $\Tr\big(\rho_0 Q_{\alpha}(\cdot )\big)$ at $\bs \in S$ is $1+\alpha \br_0 \bs$, which is even less sharp then~\eqref{e:f0}.

Finally it is worth calculating the measure of a semisphere. The ``center of mass'' of the  semisphere $S^+ (\br)$ with midpoint $\br$ ($|\br|=1$) is at $\frac{\br}{2}$, so by~\eqref{e:Qa}
\begin{equation}
Q\big( (S^+ (\br)\big) =\rho\Big( \frac{\br}{4}\Big) =
\frac{I}{2}+ \frac{\br \bsigma}{4}.
\end{equation}
Comparing it to the orthogonal projections~\eqref{e:Pxyzpm}, we see that if we are interested in a particular spin component of the state $\rho_0$, then it is better to perform a projective spin measurement than measuring $Q$. But what if we are interested in all components of the vector $\br_0$?

\subsection{Maximal qubit tomography}
%------------------------------------

In this subsection we investigate a quantum bit state estimation protocol based on the POVM $Q:\mathcal B(S)\to B(\C^2)$ introduced in the previous subsection [see equation~\eqref{e:Q}]. We find it convenient to call this protocol \emph{maximal qubit tomography}, since the whole set $S$ of pure states constitute the possible measurement outcomes.

Assume that performing $N$ independent measurements on replicas of the qubit in the same state $\rho_0$, the measurement outcomes $\{ \bn_k \}_{k=1}^N \subset S$ are obtained, where each $\bn_k$ represents a pure state on the Bloch sphere $S$, i.e., $\bn_k \in \R^3$, $|\bn_k|=1$. Using \eqref{e:f0}, the likelihood function, i.e., the probability density function on $S^N$ of obtaining this measurement statistics in the state $\rho_0 =\rho(\br_0)$ is
\begin{equation}\label{e:Lmax}
\mathcal L(\br_0) = \prod_{k=1}^N (1+\br_0 \bn_k ).
\end{equation}
Its gradient at $\br'$ is $\mathcal L(\br')\sum_{k=1}^N \frac{\bn_k}{1+\br' \bn_k}$, which yields the likelihood equation
\begin{equation}\label{e:lemax}
\sum_{k=1}^N \frac{\bn_k}{1+\br' \bn_k}=0.
\end{equation}

The second derivative of the logarithm of the likelihood function~\eqref{e:Lmax}
\begin{equation}
\frac{\partial^2}{\partial \br^2} \ln\mathcal L(\br) =
- \sum_{k=1}^N \frac{\bn_k \circ \bn_k}{(1+\br \bn_k)^2}
\end{equation}
is everywhere negative definite, so the likelihood equation~\eqref{e:lemax} \emph{has a unique} solution for $\br'$, where $\mathcal L(\br')$ is maximal. Unfortunately for a general measurement statistics $\{\bn_k\}_{k=1}^N$ this solution cannot be analytically determined. (In subsection~\ref{ss:minqubt} a similar equation~\eqref{e:lemin} was obtained, which could be explicitly solved \eqref{e:rv4} because of the symmetry of the fixed $\{\ba_k\}_{k=1}^4$ vectors.)

Instead of using the maximum likelihood estimator we introduce another obvious estimator
\begin{align}\label{e:rpmax}
\br' &=f(N) \sum_{k=1}^N \bn_k,&
&\text{with}&
f(N) &= \frac{3}{N}
\end{align}
where the (yet) unknown coefficient $f(N)$ is determined from the expectational value of $\br'$. For this aim we need two simple integrals:
\begin{align}\label{e:ints}
\int_{\bn \in S} \bn d\omega(\bn) &=0,&
\int_{\bn \in S} \bn \circ \bn d\omega(\bn) &= \frac{1}{3} I,
\end{align}
where $\omega$ is the normalized area on the unit sphere $S$. (The first integral is zero by symmetry, and the second integral can easily be calculated in spherical polar coordinates $(\vartheta,\varphi)$, where $d\omega(\vartheta,\varphi) =\frac{\sin \vartheta}{4\pi} d\vartheta d\varphi$.)

Thus, using \eqref{e:Lmax}, \eqref{e:rpmax}, and then \eqref{e:ints}, the expectational value of the estimator is
\begin{equation}
\begin{split}
\langle \br' \rangle &=
f(N) \sum_{k=1}^N \int_{\{\bn_k\} \in S^N}
\mathcal L(\br_0) \bn_k d^N\omega(\bn_k) \\
&= Nf(N) \int_{\bn \in S} \bn (1+\br_0 \bn) d\omega(\bn) =
\frac{Nf(N)}{3} \br_0,
\end{split}
\end{equation}
which means that indeed, $f(N)=\frac{3}{N}$ should be chosen to get an unbiased estimation. (The integrals decouple, and for every $k$ only the integral over $\bn_k$ gives a nontrivial result, the other integrals are $1$. The obtained result is also in accordance with~\eqref{e:rv4}.)

In order to calculate the variance we need another integral, which can be easily obtained from~\eqref{e:ints}:
\begin{equation}
\int\limits_{\bn \in S} \int\limits_{\bfm \in S}
\bn\bfm (1+\br_0 \bn) (1+\br_0\bfm ) d\omega(\bfm) d\omega(\bn)
=\frac{r_0^2}{9}.
\end{equation}
Using this, we get that
\begin{equation}
\begin{split}
\big\langle r'^2 \big\rangle &=
\frac{9}{N^2} \int_{\{\bn_k\} \in S^N} 
\Big(\sum_{k=1}^N \bn_k \Big)^2 \mathcal L(\br_0) d^N\omega(\bn_k) \\
&=\frac{9 +(N-1)r_0^2}{N},
\end{split}
\end{equation}
so, by~\eqref{e:d2}, the variance of the (unrestricted) estimator~\eqref{e:rpmax} is
\begin{equation}\label{e:Vmax}
V_{\Circle} =\frac{\big\langle r'^2 \big\rangle -\langle \br'\rangle^2}{2} 
= \frac{9-r_0^2}{2N},
\end{equation}
which exactly coincides with the one~\eqref{e:V4} obtained for the minimal qubit tomography protocol.

\section{Conclusions}
%====================

In table \ref{t:res} we summarize the results obtained for the four different qubit state estimation protocols. (The variance of the state $\rho(\br')$ is calculated from the Hilbert-Schmidt distance, and by equation~\eqref{e:d2} it is half of the variance of $\br'$.)

\begin{table*}
\begin{tabular}{|ll|c|c|}
\hline
&
\parbox{9cm}{\vspace{2pt}\textbf{Protocol}\vspace{2pt}} & 
\textbf{mean} &
\textbf{\boldmath variance of $\rho'$}
\\ 
\hhline{|==|=|=|}
$\ast$ &
\parbox{9cm}{projective spin measurements in three orthogonal directions} &
\parbox{2.2cm}{unbiased}& 
\parbox{4cm}{\vspace{2pt}
$\displaystyle \frac{9-3r_0^2}{2N}$
\vspace{2pt}}
\\ \hline
$\circledast$ &
\parbox{9cm}{POVM measurement in six orthogonal directions} &
\parbox{2.2cm}{asymptotically unbiased} &
\parbox{4cm}{\vspace{2pt}
$\displaystyle \frac{9-3r_0^2}{2N} \quad\text{for}\quad N\gg 1$
\vspace{2pt}}
\\ \hline
$\otimes$ &
\parbox{9cm}{\emph{minimal qubit tomography}\\
(POVM measurement in four tetrahedral directions)} &
\parbox{2.2cm}{unbiased} &
\parbox{4cm}{\vspace{2pt}
$\displaystyle \frac{9-r_0^2}{2N}$
\vspace{2pt}}
\\ \hline
$\Circle$ &
\parbox{9cm}{\emph{maximal qubit tomography}\\
(POVM measurement uniformly on the whole Bloch sphere)} &
\parbox{2.2cm}{unbiased} &
\parbox{4cm}{\vspace{2pt}
$\displaystyle \frac{9-r_0^2}{2N}$
\vspace{2pt}}
\\ \hline
\end{tabular}
\caption{\label{t:res}The results obtained for the different qubit tomography protocols.}
\end{table*}

It is clearly visible that all estimation schemes perform more or less equally well, although there are projective ($\ast$), discrete POVM ($\circledast$, $\otimes$), and continuous POVM measurements ($\Circle$) among them. For states close to pure states (i.e., for $r_0 \lessapprox 1$) the variance of the first two protocols with hexagonal symmetry ($\ast$ and $\circledast$) is a bit less than the variance of $\otimes$ and $\Circle$. On the other hand, the variance of the minimal and maximal qubit tomography protocol is exactly the same for all states.

Based upon these examples we may draw some general conclusions on the unusual properties of POVM measurements.

In classical Hamiltonian mechanics the phase space can be identified with the set of pure states of the system, which are simply Dirac measures concentrated on a single point of the phase space. This observation motivates to regard the set of pure states even in a finite dimensional quantum mechanical system as the quantum analog of phase space.

A widespread paradigm of quantum mechanics, based on Heisenberg's uncertainty principle, is the fact that all the classical phase space variables cannot be measured simultaneously with arbitrary high precision. Originally this was stated for \emph{projective measurements}, which are \emph{sharp} in the sense that for every possible measurement outcome there \emph{is} a state for which the specified outcome occurs with probability one. (Right after the measurement the system ``jumps'' into a state of this kind.) Coordinate and momentum operators do not commute, thus they do not possess a common projector valued measure.

On the other hand, there are generalized POVM measurements which yield results corresponding to simultaneous values of noncommuting operators! This fact is, however, not in contradiction with Heisenberg's uncertainty principle, since the result of the POVM measurement is \emph{unsharp}, i.e., for \emph{any} pure state of the system the measurement results have dispersed probability distributions. A very simple example is the POVM measurement constructed in subsection~\ref{ss:equPOVM}, which clearly demonstrates all the unusual features of generalized measurements. In summary, in contrast to projective measurements, a POVM measurement
\begin{itemize}
\item[$\ddot\smile$]
may have a continuous set of possible outcomes, even in a finite dimensional Hilbert space;
\item[$\ddot\smile$]
may yield simultaneously values for incompatible (noncommuting) observables;
\item[$\ddot\frown$]
but these simultaneous results are unsharp. (In particular, it means that repeated measurements do not give the same outcome.)
\item[$\ddot-$]
Furthermore, the information gains ($\ddot\smile$) and information losses ($\ddot\frown$) somehow compensate each other, so cleverly chosen POVM measurements yield more or less the same amount of information about the true state as projective measurements.
\item[$\ddot\smile$]
Last but not least, POVM measurements can be much better adjusted to respect continuous symmetries of the system than projective measurements.
\end{itemize}

Finally we remark that POVM measurements (or \emph{weak measurements}) are not pure mathematical constructions, they can also be experimentally realized by making an appropriate projective measurement on a composite system.

\bibliography{qubit}
\bibliographystyle{apsrev}

\end{document}